\documentstyle[aps,prl,multicol,mypsfig2,floats]{revtex}
\def\@pnumwidth{2em}

\makeatother

\begin{document}
\twocolumn
\noindent {\Large \sf Non-thermal nuclear magnetic resonance quantum computing using hyperpolarized Xenon}

\noindent {\large \sf Anne S. Verhulst}\\
Solid State  and Photonics Laboratory, Stanford University, Stanford, CA 
94305\\
IBM Almaden Research Center, San Jose, CA 95120 \\
\noindent {\large \sf Oskar Liivak, Mark H. Sherwood}\\
IBM Almaden Research Center, San Jose, CA 95120 \\
\noindent {\large \sf Hans-Martin Vieth}\\
Institute of Experimental Physics, Free University of Berlin, Berlin, 
Germany\\
\noindent{\large \sf Isaac L. Chuang}\\
IBM Almaden Research Center, San Jose, CA 95120 \\
Media Laboratory, Massachussetts Institute of Technology,
Cambridge, MA 02139 \\


\def\>{\rangle}
\def\be{\begin{equation}}
\def\ee{\end{equation}}
\def\bea{\begin{eqnarray}}
\def\eea{\end{eqnarray}}
\newcommand{\ket}[1]{\mbox{$|#1\rangle$}}
\newcommand{\bra}[1]{\mbox{$\langle #1|$}}
\newcommand{\mypsfig}[2]{\psfig{file3D#1,#2}}



\vspace*{0.5cm}
\begin{abstract}
\vspace*{-1cm}

Current experiments in liquid-state nuclear magnetic resonance
quantum computing are limited by low initial polarization.
To address this problem, we have investigated the use of optical pumping techniques to enhance the polarization of a 2-qubit NMR quantum computer ($^{13}$C and $^1$H in $^{13}$CHCl$_3$).  To efficiently use the increased polarization, we have generalized the procedure for effective pure state preparation.  With this new, more flexible scheme, an effective pure state was prepared with
polarization-enhancement of a factor of 10
compared to the thermal state.  An implementation of Grover's quantum search algorithm was demonstrated using this new technique.\\
\end{abstract}


Intensive experimental efforts have been made to implement quantum 
computing
since Shor~\cite{shor_94} and Grover~\cite{grover_97} developed their
respective algorithms. So far, the only experimental demonstrations of
quantum computing algorithms have been based on nuclear magnetic
resonance (NMRQC)~\cite{recent-nmrqc-achievements}. However, the low
polarization of high-temperature nuclear spin systems is a major
limitation of this approach. To cope with the corresponding highly
mixed  spin state, schemes have been developed which produce an
effective pure state~\cite{labeling,knill_98}.
Unfortunately, all of these techniques require exponential resources
to implement and improved scalable procedures for fully polarizing an embedded spin system~\cite{cooling} are not practical with the very low 
initial
polarizations found in NMRQC.
In addition to the experimental challenges, recent theoretical work has
called into question the boundary between classical and quantum
computing in the case of highly mixed systems~\cite{schack_caves_99}.
A crucial parameter in determining where this boundary lies is the initial polarization of the quantum system.
It is therefore highly relevant to explore existing techniques
used to enhance polarization for NMR and to investigate
their application to quantum computing~\cite{hubler_00}.

One recently developed technique is based on laser-polarized Xenon.
It is known that the high electron polarization in
optically pumped alkali metals like rubidium can be transferred to
the nuclear spins of noble gases such as $^3$He or $^{129}$Xe via
a spin-exchange mechanism~\cite{opticalpumping}. More recently, it
was found that hyperpolarized liquid $^{129}$Xe could be used to
enhance the polarization of nuclei of other molecules in solution
via the spin polarization-induced nuclear Overhauser
effect (SPINOE)~\cite{spinoe,song}.
The focus of our work is the coupling of optical pumping with NMRQC.
We have developed a temporal labeling scheme which is more flexible than 
the existing ones and which allows efficient use of the polarization 
enhancements.
To validate this technique, we have created an effective pure state
with an order of magnitude polarization enhancement. The flexibility of 
our scheme
is further illustrated by applying this approach
to  perform Grover's search algorithm. Polarization-enhancements up to a 
factor of 7 are achieved.

\begin{figure}[htbp]
\begin{center}
\mbox{\psfig{file=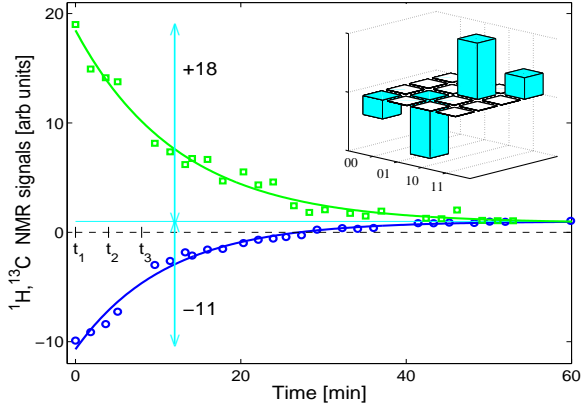,width=3.3in}} 
\end{center}
\caption{Time evolution of the $^{13}$C ($\Box$) and $^1$H ($\circ$)
signals of $^{13}$CHCl$_3$ after being dissolved in hyperpolarized
$^{129}$Xe for 4 min at -40$^o$C. The initial enhancement
is +18 for $^{13}$C and -11 for $^1$H.
The inset shows the deviation density matrix of
the \{$^1$H$^{13}$C\} spin state at time $t_1$ (the elements on the 
diagonal
are experimental data - the off-diagonal elements are known to be 
zero).}
\label{fig:enhancements}
\end{figure}

In a typical optical pumping experiment, we start by pressurizing
a glass cell (pyrex, 200 cc) containing solid Rb
to 3.5 Atm with a mixture of ultrapure gases: $12\%$ natural
abundance Xe, $2\%$ N$_2$ and $86\%$ He. After the cell is heated
to $110^{\circ}$C to produce a saturated Rb vapor, a 100 W diode array
laser (Optopower), tuned to the D1 transition of Rb, is switched on.  This circularly polarized light irradiates the gas mixture and
polarizes the Rb electrons. After 20 min. of Rb-Xe spin exchange,
the gas mixture is passed through a cold trap which condenses the
highly polarized Xe. The Xe is then frozen (77K) in a high-pressure
NMR tube (New Era) containing 20\textup{$\mu$}l of
degassed $^{13}$CHCl$_3$. The sample tube is moved into a 2.1T magnet and 
the
temperature regulated to $-40^{\circ}$C so that $^{13}$CHCl$_3$
and Xe liquefy and mix (Xe:$^{13}$CHCl$_3$ molar ratio = 5:1 to 8:1).
The hyperpolarized liquid $^{129}$Xe enhances
the polarization of  $^1$H and $^{13}$C in $^{13}$CHCl$_3$
via the SPINOE effect.  Results are monitored using a homebuilt
spectrometer (see Fig.~\ref{fig:enhancements}).
The data show that the enhancements are large in the
Xe-$^{13}$CHCl$_3$ mixture. We have achieved polarizations over a
factor of 10 greater than the thermal equilibrium value for
both $^1$H and $^{13}$C.

The deviation density matrix
($\rho = qI + \rho_{dev}$)~\cite{labeling,knill_98} of
the \{$^1$H$^{13}$C\} spin state at time $t_1$ is shown in the inset
to Fig.~\ref{fig:enhancements}. The data were obtained by
observing $^1$H and $^{13}$C simultaneously after sending two small
tip angle pulses, providing full information about the diagonal elements
of the deviation density matrix~\cite{chuang_bulk_quantum}.
Even though the polarizations of $^{13}$C and $^1$H are larger than
in the thermal case, the spin state is still highly mixed.

Labeling schemes are applied to create an effective pure state from
a mixed state
($\rho_{\mathit{eff}\, pure}
= q_1I + q_2\,\rho_{pure}$)~\cite{labeling,knill_98}.
Temporal labeling schemes cyclicly permute ($P_i$) the populations
of all states but the ground state. The resulting $2^n-1$
density matrices -- with n the number of quantum bits -- are 
summed
to create an
effective pure state ($\rho_{\mathit{eff}\, pure} =
  \sum_{i=1}^{2^n-1} P_i\rho_{init}P_i^{\dagger}$).
If the initial state is reproducible and its density matrix diagonal
in the computational basis, the optimal choice of ground state (in terms 
of
signal-to-noise) and the corresponding set of permutations ($P_i$)
can be determined~\cite{knill_98}.
The polarization-enhanced initial states are known to be diagonal
(cross-relaxation, the mechanism underlying SPINOE,
is not coherent). However, there are small changes in initial 
enhancement
from experiment to experiment and therefore the initial density matrices
are not all identical for each of the $2^n-1$ cyclic permutation
experiments. Therefore, we generalized the existing temporal labeling
scheme by introducing weights $w_i$ in the summation and allowing $\rho_{init}$ to vary from experiment to experiment:
  \be  \rho_{\mathit{eff}\, pure}
                 = \sum_{i=1}^{2^n-1} =
w_i(P_i\rho_{init,i}P_i^{\dagger}) \, .
  \label{eq:templab}
  \ee
Provided one can determine $\rho_{init,i}$ experimentally, the weights  can be calculated since
the matrix representation of Eq.\ref{eq:templab}
yields a set of $2^n-1$ linear equations in the $2^n-1$ variables
$w_i$.

\begin{figure}[htbp]
\begin{center}
\mbox{\psfig{file=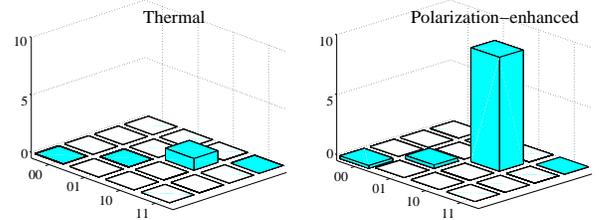,width=3.3in}} 
\end{center}
\caption{Deviation density matrix of the thermal and
polarization-enhanced effective pure state (the elements on the
diagonal are experimental data -- the off-diagonal
elements are known to be zero). With our
generalized temporal labeling scheme and using one optically pumped
sample per permutation experiment, we have created an
effective pure state with a factor of 9.5 polarization-enhancement.}
\label{fig:effpure}
\end{figure}

We applied this generalized temporal labeling procedure to an
n=2-qubit quantum computer with polarization-enhanced initial states.
The density matrix $\rho_{init,i}$ is inferred by adding a
probing experiment a short time $r_1$ before the permutation experiment. 
The probing experiment consists of two
simultaneous RF pulses with small tip angle ($10^\circ$ to $20^\circ$)
on both spins leading to results like those shown in the inset to
Fig.~\ref{fig:enhancements}. The time $r_1$
is calculated to be long enough so that the system has returned to
its quasi-equilibrium enhanced state, yet short enough
($r_1$ = 25 sec $\ll T_{1,Xe}$ = 15 min) so that the density matrix
inferred from the probing experiment closely approximates
the state $\rho_{init,i}$ at the start of the permutation experiment.
The full effective pure state preparation thus consists of the 
preparation
of three ($2^n-1$) optically pumped samples. On each sample,
a permutation experiment is performed, preceded by a probing
experiment to gain information on $\rho_{init,i}$. Then 
Eq.~\ref{eq:templab}
is solved for $w_i$ and the enhancement of
$\rho_{\mathit{eff}\, pure}$ over the effective pure state
resulting from experiments at thermal equilibrium is
determined (see Fig.~\ref{fig:effpure}).
An effective pure state with a factor of 9.5 polarization
enhancement was created.

We further illustrate the flexibility of our generalized temporal
labeling scheme by performing multiple experiments using one
optically pumped sample. In terms of polarization, this
can be done since  Xe, which has a very long relaxation
time ($T_{1,Xe}$ = 15 min), is a quasi-continuous source of
high polarization. After each experiment, which lasts less than a 
second, 
the time for $^{13}$CHCl$_3$ to return to a quasi-equilibrium enhanced
state is 2 min ($5T_{1C,H}$). Therefore, multiple experiments can
be performed before the Xenon polarization relaxes significantly.
Thus, instead of preparing 3 separate optically pumped samples,
only one sample is prepared on which 3 permutation experiments
are performed (see typical starting times $t_1$, $t_2$ and $t_3$
in Fig.~\ref{fig:enhancements}).
The $\rho_{init,i}$ are now significantly different since the 
polarization
is decreasing, but a set of weights $w_i$ can still be calculated from
Eq.~\ref{eq:templab} to create an effective pure state.
It can even be proven that as long as the set of
{$\rho_{init,i}$} are diagonal matrices, Eq.~\ref{eq:templab} represents 
the temporal labeling procedure with optimal 
signal-to-noise~\cite{general}.
This variant of the labeling scheme was used to create the input states
of a quantum computation.

\begin{figure}[!htb]
\begin{center}
\mbox{\psfig{file=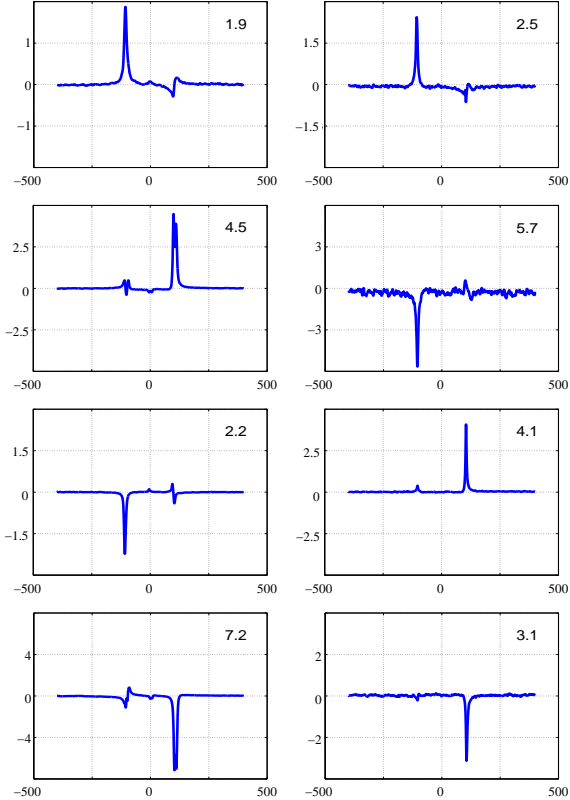,width=3.3in}} 
\end{center}
\caption{Spectral readout of the four cases of the 2-qubit
Grover search using the polarization-enhanced initial state
of $^{13}$CHCl$_3$. The plots show the real part of the $^1$H
(left) and $^{13}$C (right) spectra, with NMR lines
at $\pm \frac{1}{2}J_{CH}$  (in Hz relative to
$\nu_H$ and $\nu_C$). Positive (negative) peaks correspond to the state
$\ket{0}$ ($\ket{1}$). The vertical scale is arbitrary.  From top to bottom the marked element is $x_o$ = 00, 01, 10 or 11
and as the spectra show the Grover algorithm outputs exactly this
element. The data are obtained by using our generalized
temporal labeling scheme and by performing multiple experiments with
one optically pumped sample. Each instance of the Grover
algorithm is executed twice, reading out first the $^{1}$H and then
the $^{13}$C spectra. The numbers on the plots represent
the polarization-enhancements achieved. They vary between 2 and 7.}
\label{fig:grover}
\end{figure}

We demonstrated this technique by implementing Grover's search algorithm; unlike the Deutsch-Josza algorithm~\cite{D-J}, this requires a pure input state
to produce a meaningful result and therefore is more demanding.
The goal of the 2 qubit Grover algorithm is to identify an element $x_o$ among
four possible elements $x_i$ by querying an oracle function $f(x)$ for
which $f(x_o) = 1$ while $f(x_i \neq x_o) = 0$. The four elements
$x_i$ are represented by the spin states
$\ket{00},\ket{01},\ket{10},\ket{11}$ of \{$^1$H$^{13}$C\}
($\ket{0}$ and $\ket{1}$ correspond to nuclear spin state up and down
respectively).
Classically this search would take an average of
2.25 attempts, while one query is sufficient using the Grover
algorithm~\cite{chuang_98,Jones_grover}. The output of Grover's search
algorithm is the state $\ket{x_o}$.
Using our generalized labeling scheme, the experimental implementation
starts with a probing experiment to gain information about $\rho_{init,i}$. 
A time $r_1$ later the quantum computation experiment is performed, 
which is a concatenation of the pulses of the cyclic permutation $P_i$ and
the pulse sequence representing the actual quantum algorithm
(same protocol used in ~\cite{chuang_98}). The resulting
$^1$H and $^{13}$C readout spectra are compared with the thermal spectra
(Fig.~\ref{fig:grover}). We succesfully
implemented the four possible cases $x_o$ = 00, 01, 10 or 11 of
Grover's quantum algorithm with polarization enhancements
as large as a factor of 7.

In conclusion, we have demonstrated that with optical pumping and
using our generalized temporal labeling scheme, a
polarization-enhanced effective pure state can be
produced and a quantum computation can be performed. The polarization
enhancements of more than a factor of 10 for both $^{13}$C and $^1$H
in $^{13}$CHCl$_3$ are comparable to other published results using
hyperpolarized Xe and SPINOE~\cite{song}.
Even though large scale NMR quantum computers are not yet within reach,
the first steps for the necesary polarization-enhancement have now been
taken. Further increases in polarization can be achieved by using
isotopically  pure $^{129}$Xe, which increases the polarization by a
factor of 3 to 4, by improvements in the design of the pumping 
apparatus,
as well as by the screening of other candidate quantum computing 
molecules.
Moreover, the effective control over a large range of
initial polarizations could allow NMRQC to explore the fundamental
divergence between quantum computing and classical computing.

The authors would like to acknowledge D. Pomerantz, S. English and J.-P. 
Strachan for help, L. Vandersypen, L. Kaiser, C. Yannoni and X. Zhou for 
useful discussions, C. Yannoni and R. Macfarlane for use of equipment 
and J. Harris and W. Risk for their support. A. Verhulst gratefully
acknowledges a Francqui Fellowship of the Belgian American Educational
Foundation. This work was supported by DARPA under the NMRQC initiative.



\begin{thebibliography}{16}
\bibitem{general}
Every permutation operation $P_i$ will now depend on the given $\rho_{init,i}$ and on the desired ground state (for n=2, $\ket{10}$ might be the preferred ground state if one of the spins is negatively enhanced). The weights are determined in a similar way as described in the letter.
\end{thebibliography}
\end{document}